\documentclass[aps,prl,twocolumn,showpacs,superscriptaddress,groupedaddress,preprintnumbers,amsmath,amssymb]{revtex4}

\usepackage{graphicx}
\usepackage{dcolumn}
\usepackage{bm}
\usepackage[amssymb]{SIunits}
\usepackage{color}
\usepackage{ulem}
\usepackage{textcomp}

\begin{document}

\preprint{APS/123-QED}

\title{All-optical formation of coherent dark states of silicon-vacancy spins in diamond}

\author{Benjamin Pingault}
\thanks{These authors contributed equally to this work.\\}
\affiliation{Cavendish Laboratory, University of Cambridge, JJ Thomson Ave, Cambridge CB3 0HE, UK}

\author{Jonas N. Becker}
\thanks{These authors contributed equally to this work.\\}
\affiliation{Fachrichtung 7.2 (Experimentalphysik), Universit\"{a}t des Saarlandes, Campus E2.6, 66123 Saarbr\"{u}cken, Germany}

\author{Carsten H. H. Schulte}
\affiliation{Cavendish Laboratory, University of Cambridge, JJ Thomson Ave, Cambridge CB3 0HE, UK}

\author{Carsten Arend}
\affiliation{Fachrichtung 7.2 (Experimentalphysik), Universit\"{a}t des Saarlandes, Campus E2.6, 66123 Saarbr\"{u}cken, Germany}

\author{Christian Hepp}
\affiliation{Cavendish Laboratory, University of Cambridge, JJ Thomson Ave, Cambridge CB3 0HE, UK}

\author{Tillmann Godde}
\author{Alexander I. Tartakovskii}
\affiliation{Department of Physics and Astronomy, University of Sheffield, Sheffield S3 7RH, UK}

\author{Matthew Markham}
\affiliation{Element Six Ltd., Global Innovation Centre, Fermi Avenue, Harwell Oxford, Didcot, OX11 0QR, UK}

\author{Christoph Becher}
\thanks{Electronic address: christoph.becher@physik.uni-saarland.de, ma424@cam.ac.uk\\}
\affiliation{Fachrichtung 7.2 (Experimentalphysik), Universit\"{a}t des Saarlandes, Campus E2.6, 66123 Saarbr\"{u}cken, Germany}
\email{christoph.becher@physik.uni-saarland.de}

\author{Mete Atat\"{u}re}
\thanks{Electronic address: christoph.becher@physik.uni-saarland.de, ma424@cam.ac.uk\\}
\affiliation{Cavendish Laboratory, University of Cambridge, JJ Thomson Ave, Cambridge CB3 0HE, UK}
\email{ma424@cam.ac.uk}

\date{\today}

\begin{abstract}
Spin impurities in diamond can be versatile tools for a wide range of solid-state-based quantum technologies, but finding spin impurities which offer sufficient quality in both photonic and spin properties remains a challenge for this pursuit. The silicon-vacancy center has recently attracted a lot of interest due to its spin-accessible optical transitions and the quality of its optical spectrum. Complementing these properties, spin coherence is essential for the suitability of this center as a spin-photon quantum interface. Here, we report all-optical generation of coherent superpositions of spin states in the ground state of a negatively charged silicon-vacancy center using coherent population trapping. Our measurements reveal a characteristic spin coherence time, $\textrm{T}_{2}^{*}$, exceeding 250 nanoseconds at \unit{4}{K}. We further investigate the role of phonon-mediated coupling between orbital states as a source of irreversible decoherence. Our results indicate the feasibility of all-optical coherent control of silicon-vacancy spins using ultrafast laser pulses. 
\end{abstract}

\maketitle

Confined impurity spins in spin-free materials such as diamond and silicon offer a multitude of opportunities ranging from fundamental studies of engineered mesoscopic spin system dynamics to potential applications emerging from quantum control. A fundamental advantage of diamond-based impurities, known as color centers, is that they can be optically active in the conveniently detectable visible to near infrared region of the spectrum \cite{Zaitsev2001,Prawer2014}. Of these, the nitrogen-vacancy center (NV) remains the most studied one \cite{Clark1971,Acosta2013,Doherty2013,Childress2013}. Sharing its desirable and undesirable properties alike, a handful of other impurities have recently been investigated \cite{Aharonovich2014}. These investigations reveal, for the NV center, the presence of crystal-field splitting in the ground state manifold allowing for feasible microwave control \cite{Maze2011,Doherty2011,Gruber1997,Dolde2014}. However,  the unfavorable, but dominant, emission into phonon sidebands also occurs in these centers. Contemporary research efforts focus on two parallel approaches: Amplifying the zero-phonon emission by coupling selectively to an optical mode of a cavity \cite{Aharonovich2014, Li2014, Aharonovich2011} and  investigating alternative color centers with sufficiently small phonon sideband contribution to the full optical spectrum \cite{Aharonovich2014,Aharonovich2011}.

The negatively charged silicon-vacancy (SiV$^{\text{-}}$) center is a particularly interesting justification to pursue the latter of the two approaches: The optical transitions coupling the excited and the ground state manifolds are predominantly into the zero-phonon line \cite{Neu2011a}, which can be further enhanced by making use of the ongoing progress in diamond-based optical cavity nanostructures \cite{Riedrich-Moeller2014}. Also, the impressively small variation in the emission spectrum among multiple SiV$^{\text{-}}$ centers in a clean diamond matrix \cite{Sipahigil2014,Rogers2014} deems them desirable for coupling multiple spins via a common photonic mode. In parallel, recent demonstrations of the direct optical access to the spin degrees of freedom of single SiV$^{\text{-}}$ centers \cite{Mueller2014} offer the exciting possibility to employ full quantum control relying only on optical fields \cite{Santori2006,Yale2013}, which can bring the speed-up advantage of optics over control techniques in the microwave regime. However, there are a number of open questions that need to be answered before such steps forward can be taken. Arguably, the most pressing challenge is to determine the coherence time of the SiV$^{\text{-}}$ spin in the ground state in the presence of the potentially detrimental coexistence of the spin and orbital degrees of freedom. In this Letter, we achieve coherent population trapping (CPT) between Zeeman-split states as a means to generate a coherent superposition, i.e. coherent dark state, of a single SiV$^{\text{-}}$ center spin. We report a spin coherence time ($\textrm{T}_{2}^{*}$) lower bound of 250 ns - more than two orders of magnitude longer than the optical transition timescale \cite{Rogers2014,Sternschulte1994}. We first identify the operational conditions for generating the $\Lambda$ system required for CPT by controlling the angle of the applied magnetic field. We further investigate the role of phonons as a source of decoherence within the ground state by tuning the spin states across an avoided crossing, where spin orthogonality is relaxed.

\begin{figure*}[t]
\centering
\includegraphics[width=1\textwidth]{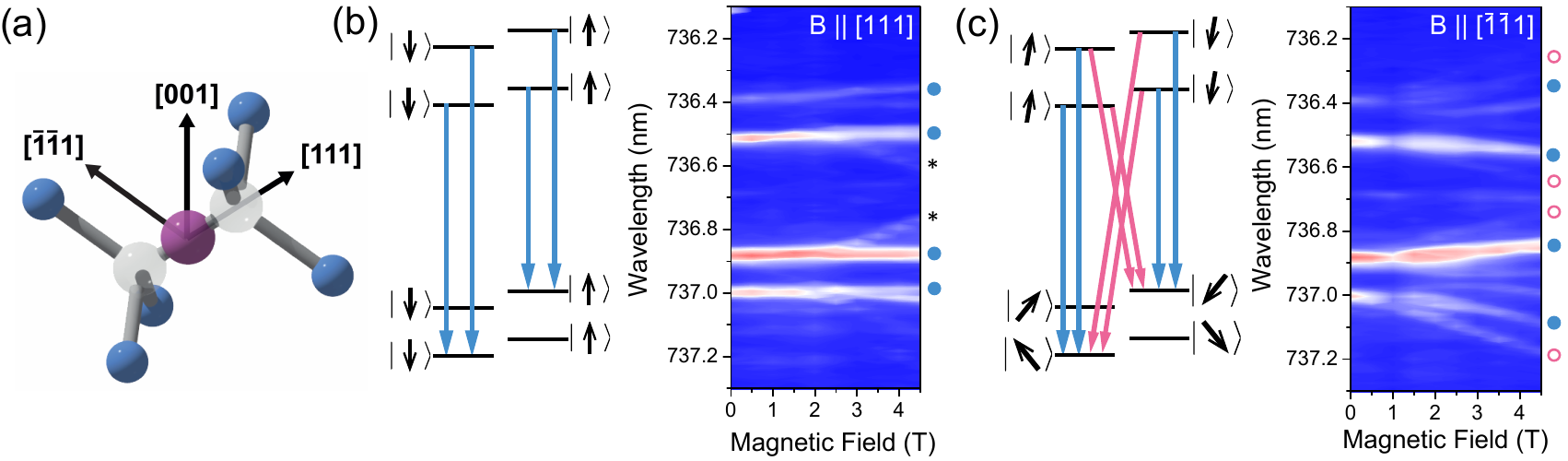}\\
\caption{(Color online) (a) Atomic structure of the SiV$^{\text{-}}$ color center, consisting of a Si impurity (purple) situated on an interstitial position along the [111] bond axis and surrounded by a split-vacancy (light grey) and the next-neighbor carbon atoms (blue). (b, left) Resulting energy levels and spin projections for magnetic fields applied along [111]. The level scheme shown here is simplified (a detailed scheme can be found in \cite{Supplement}). Optical transitions (blue arrows) are allowed between levels of the same spin state and the most visible ones are marked by blue dots in the magnetic field-dependent non-resonant fluorescence spectrum at \unit{4}{K} (excitation at \unit{660}{nm}) (b, right). Applying the magnetic field along the $[\bar{1}\bar{1}1]$ direction on the same SiV$^{\text{-}}$ center, transverse field components lead to a finite spin overlap for all ground and excited states (c, left), resulting in additional optical transitions (pink arrows), observed in the field dependent spectrum (c, right). The spin labels refer to a Bloch vector representation, as explained in \cite{Supplement}.}
\end{figure*}

We investigate two samples, an electronic grade (001)-oriented CVD diamond used for magnetic-field orientation measurements and a (111)-oriented type IIa high-pressure-high-temperature (HPHT) diamond used for spin coherence measurements. SiV$^{\text{-}}$ centers are generated by $^{28}$Si implantation followed by thermal annealing. To enhance the optical excitation and collection efficiencies, arrays of solid immersion lenses (SILs) are fabricated on the surfaces of both samples using a focused ion beam (FIB) (for further details, see \cite{Supplement}). All our experiments are carried out at \unit{4}{K}.

An SiV$^{\text{-}}$ center is formed by a substitutional silicon atom and a vacancy replacing two neighboring carbon atoms in the diamond matrix along the $\langle111\rangle$ axes. The silicon atom relaxes to the interstitial lattice site to form an inversion-symmetric split-vacancy structure [see Fig.\,1(a)] \cite{Goss1996}. The spin-orbit coupling dictates an inherent quantization axis for the spin degree of freedom aligned with the SiV$^{\text{-}}$ symmetry axis in both the ground and the excited state manifolds \cite{Hepp2014}. Figure 1(b) displays the fluorescence spectrum from a single SiV$^{\text{-}}$ center in the (001) sample under non-resonant excitation with a magnetic field applied along this inherent $[111]$ quantization axis. The dominant optical transitions (four of which marked by blue filled circles) conserve the spin state, as illustrated in the accompanying energy level scheme. Weaker transitions, identified by asterisks, arise due to a slight mismatch between the symmetry axis and the direction of the applied magnetic field. This serves to reveal the importance of the magnetic field orientation for optical transition rules in the SiV$^{\text{-}}$ center level scheme. 

A magnetic field, applied at a finite angle to the [111] direction, constitutes an external quantization axis which competes with the SiV$^{\text{-}}$ center's internal counterpart. The ground and excited state manifolds experience different strength of spin-orbit interaction \cite{Hepp2014}. Consequently, this configuration gives rise to different effective quantization axes between the two manifolds. The net angle between these resultant quantization axes, in turn, determines the optical selection rules for the fluorescence spectrum. Figure 1(c) presents the same measurement as Fig.\,1(b), but for a $[\bar{1}\bar{1}1]$-oriented magnetic field. The spin selectivity of the optical transitions no longer holds and new optical transitions arise as the strength of the magnetic field increases. Four of these additional transitions are indicated by pink open circles and pink arrows. In summary, the fully aligned magnetic field case [Fig.\,1(b)] yields cycling transitions, whereas a magnetic field at an angle allows for typical $\Lambda$ schemes, where two orthogonal spin ground states can have finite transition matrix elements to the same excited state. This provides the desired configuration for all-optical manipulation of the SiV$^{\text{-}}$ spin via this shared excited state. 

Figure 2 illustrates the detection strategy and characterization of the SiV$^{\text{-}}$ center used for coherent population trapping. We start by identifying a bright SiV$^{\text{-}}$ center within the SIL array of the (111) sample. Superimposed images of electron and fluorescence microscopy scans for the same area of the sample, as shown in Fig.\,2(a), demonstrate an example of enhanced SiV$^{\text{-}}$ fluorescence under one of the SILs. Figure 2(b) shows the detection concept for all single- and multi-laser resonant excitation experiments, where the signal is obtained by measuring the integrated fluorescence from the transitions in the shaded area as a function of the excitation laser frequency. Non-radiative decay into the lower orbital branch of the excited state followed by fluorescence allows us to monitor excited-state population directly with no residual laser contribution \cite{Mueller2014}. In order to allow $\Lambda$ schemes, the angle between the external magnetic field and the SiV$^{\text{-}}$ center axis is set to 109.4\degree, i.e.\ the angle between $[\bar{1}\bar{1}1]$ and [111] directions [see Fig.\,1(a)]. Pulsed intensity-correlation measurements performed on the selected SIL suggest the presence of two individual SiV$^{\text{-}}$ centers with strong spatial and spectral overlap \cite{Supplement}. Single-laser resonant excitation of the D1 transition under 3\,T magnetic field resolves the resonances of the two centers spectrally owing to their slightly differing (2\%) strain, as shown in Fig.\,2(c). This slight variation in the strain tensor between the two centers is used to address each SiV$^{\text{-}}$ selectively. The following experiments are performed using emitter 1 in Fig.\,2(c).

\begin{figure}[t]
\centering
\includegraphics[width=0.49\textwidth]{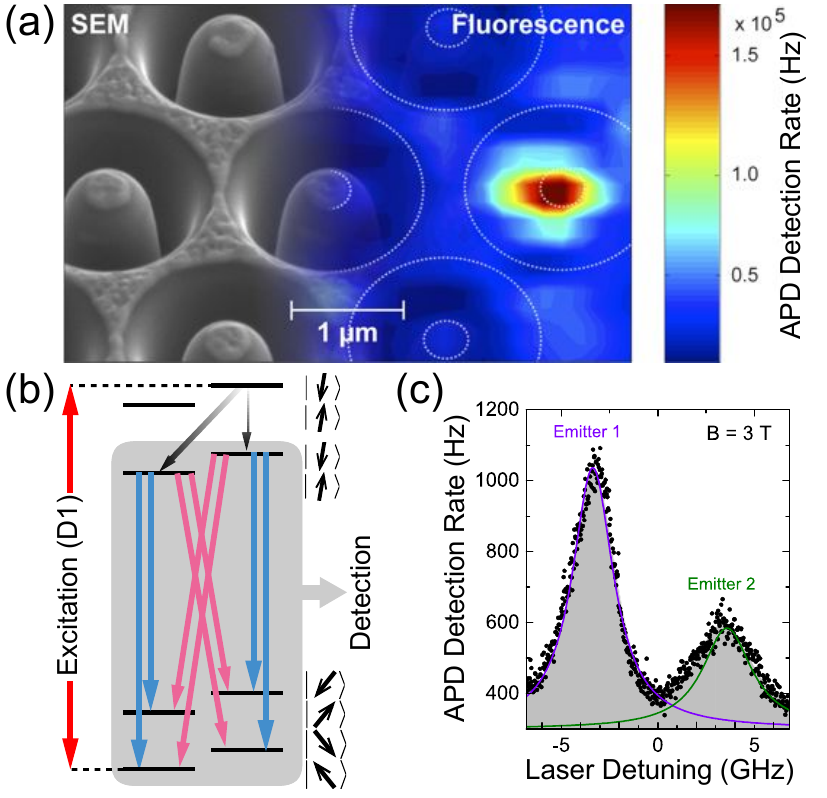}\\
\caption{(Color online) (a) Scanning electron microscope image of the solid immersion lens array on the HPHT sample, superimposed by a corresponding fluorescence image (exc. 690\,nm, det. 730-750\,nm). (b) Optical excitation is performed resonantly to the highest energy exited state (transition D1, thick red arrow), from where a relaxation to lower excited states (black arrows) occurs, followed by an optical decay to the ground state (red/blue arrows). The emitted fluorescence photons are detected as a function of the excitation frequency. (c) At B = 3\,T, resonant excitation reveals the presence of two SiV$^{\text{-}}$ emitters, spectrally separated by approximately 8\,GHz.}
\end{figure}

\begin{figure}[b]
\centering
\includegraphics[width=0.49\textwidth]{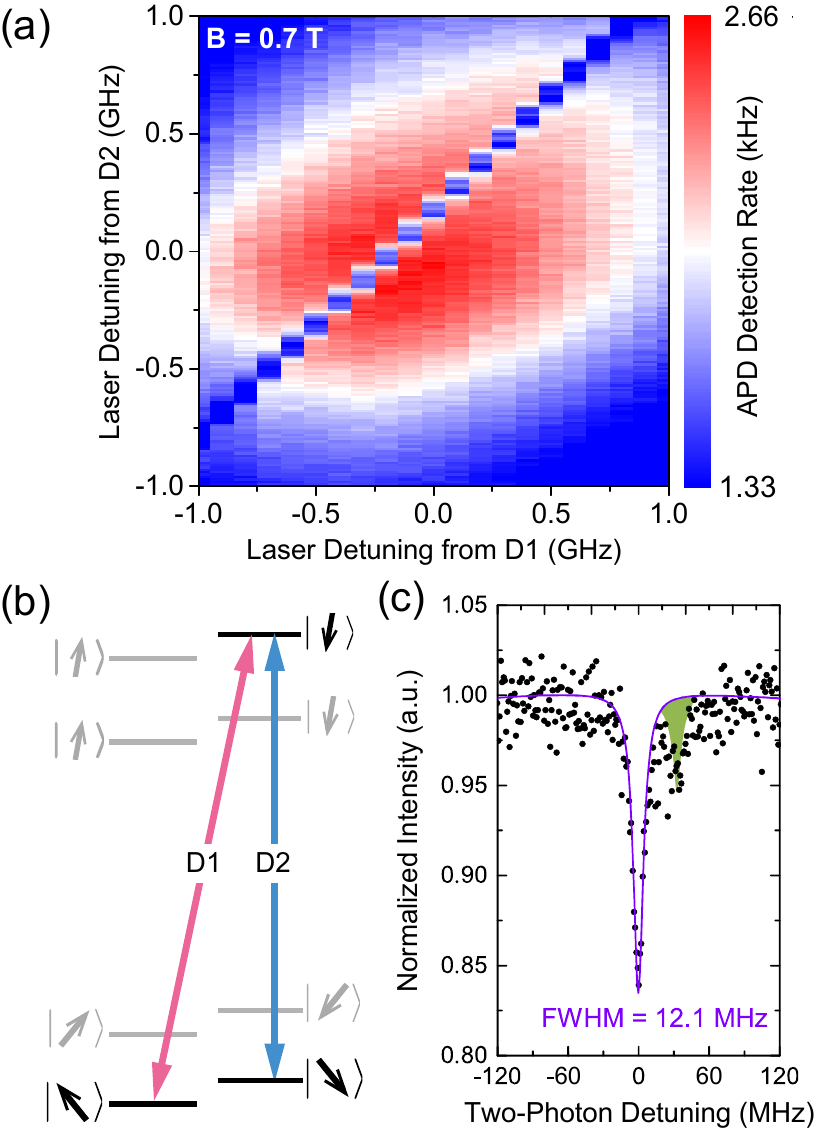}\\
\caption{(Color online) (a) CPT scan: SiV$^{\text{-}}$ fluorescence intensity recorded as the frequency of the laser resonant with transition D2 is scanned and the laser resonant with transition D1 is fixed at a given frequency. Laser powers are equal to approximately four times and seven times the saturation powers for transitions D2 and D1 respectively. (b) Relevant level structure and the driven transitions D1 and D2. (c) CPT scan at low driving power (0.33\,\textmu{W} each, corresponding to the saturation power for the D1 transition and half the saturation power for D2) yielding a dip full width at half maximum of 12.1 MHz. The purple line corresponds to a fit using a model based on optical Bloch equations \cite{Supplement} and giving a decoherence rate between the two ground states of \unit{4.0 \pm 0.2}{MHz}. The green filled curve at slightly higher frequency is the CPT dip of the second SiV$^{\text{-}}$ center.}
\end{figure}

If the two transitions of a $\Lambda$ system are driven simultaneously, the spin is optically pumped into a coherent superposition of the two ground states (dark state) determined by the two optical fields; a technique known as CPT \cite{Fleischhauer2005}. As a consequence of destructive quantum interference, optical excitation to the shared excited state and, consequently any fluorescence originating from this state, is suppressed. The reduction of the integrated fluorescence, i.e. the CPT dip, is strongly dependent on the coherence between the two ground states and its spectral width allows direct measurement of the coherence timescale of the ground state \cite{Fleischhauer2005}. Figure 3(a) presents a two dimensional CPT scan for the SiV$^{\text{-}}$ at 0.7\,T magnetic field as a function of the optical frequencies of the two lasers driving the D1 and D2 transitions selectively, as illustrated in Fig.\,3(b). The two ground states addressed originate from the same orbital branch and have ortho\-gonal spin projections \cite{Supplement}. The manifestation of CPT is evident as a significant drop of the fluorescence intensity at two-photon resonance ($\delta_{\textrm L1}-\delta_{\textrm L2} = \Delta$, where $\delta_{\textrm L1}$ and $\delta_{\textrm L2}$ denote the laser detunings from the D1 and D2 transitions and $\Delta$ is the frequency difference between the two states). Figure 3(c) presents the CPT dip obtained by scanning the frequency of the laser driving the D2 transition, while keeping the D1 excitation fixed on resonance. In order to extract the ground state coherence time both lasers are kept at sufficiently low excitation powers (equal to saturation power for the D1 transition and half the saturation power for D2) in order to minimize power broadening effects in the CPT dip. Using a Lorentzian fit, the full width at half maximum of the CPT dip under these conditions is 12.1\,MHz. A theoretical model, based on optical Bloch equations which include the excited state dephasing, residual power broadening and the mutual coherence of the lasers, is used to fit the data with the decoherence rate between the two ground states as a free parameter. This provides an upper bound of \unit{4.0 \pm 0.2}{MHz} for the ground states coherence contribution to this width and therefore a measure of the SiV$^{\text{-}}$ ground-state coherence time exceeding 250\,ns. 

The observed coherence time is more than two orders of magnitude longer than the timescale for thermalization, which typically takes place within a nanosecond \cite{Mueller2014}. Hence, we suggest that it is the spin that dictates the decoherence mechanism for the ground states, as the phonon-induced thermalization for ground states of opposite spin is quenched. To support this argument, we take advantage of the presence of an avoided crossing at 3.5\,T between two of the ground states. By sweeping the magnetic field over the region of the avoided crossing, we relax the spin state orthogonality, thus progressively allowing for phonon-mediated decoherence of the dark state. Figure 4(a) depicts the evolution of the spin for the ground states coupled by CPT, as the magnetic field is varied over the avoided crossing. From 0 to 3.5\,T, the dark state is generated between states $|1\rangle$ and $|2\rangle$, while above 3.5\,T the dark state is generated between $|1\rangle$ and $|3\rangle$, as illustrated by the red and blue ribbons respectively \cite{Supplement}.

Figure 4(b) shows the linewidth of the CPT dip as a function of the magnetic field ($|1\rangle$ - $|2\rangle$ as red filled circles, $|1\rangle$ - $|3\rangle$ as blue filled circles). This width is proportional to the decoherence rate between the two driven states, on top of a constant power broadening due to the lasers \cite{Gateva2005}. The dip width increases rapidly when approaching the avoided crossing, and reaches minimum values for both low and high field limits. We calculate the spin-overlap between the two driven states and display it as dashed gray lines Fig.\,4(b) \cite{Supplement,ScalFact}. This simple approach already describes the observed trend, emphasizing the central role of the spin orthogonality of the two ground states for decoherence. The spin overlap is multiplied by a Boltzmann factor (red and blue solid lines \cite{Supplement,ScalFact}), thus taking into account the thermal activation of phonons between the addressed ground states, as their energy difference increases with increasing magnetic field. A detailed description of the phonon-mediated mechanism, such as phonon scattering or dynamic Jahn-Teller distortion, can be identified after a temperature-dependent investigation is performed. The agreement between our simple model and the experimental data confirms the hypothesis that the ground state coherence time of \unit{250}{ns} measured away from the avoided crossing corresponds to the coherence of the spin in the driven ground states. This spin coherence time is identified as the free induction decay time ($T_{2}^{*}$). It is worth noting that the sample employed for the CPT measurement shows evidence for a considerable concentration of substitutional nitrogen [N$_{S}^{0}$] \cite{Supplement}, which is known to be the main limitation for $T_{2}^{*}$ of the nitrogen vacancy spin \cite{Jelezko2004}. The same mechanism is likely to affect the measured $T_{2}^{*}$ for the SiV$^{\text{-}}$ center. Consequently,  this coherence time can be extended using all-optical pulsed protocols analogous to the dynamical decoupling techniques commonly applied to the NV center \cite{DeLange2010}. 

\begin{figure}[htb]
\centering
\includegraphics[width=0.49\textwidth]{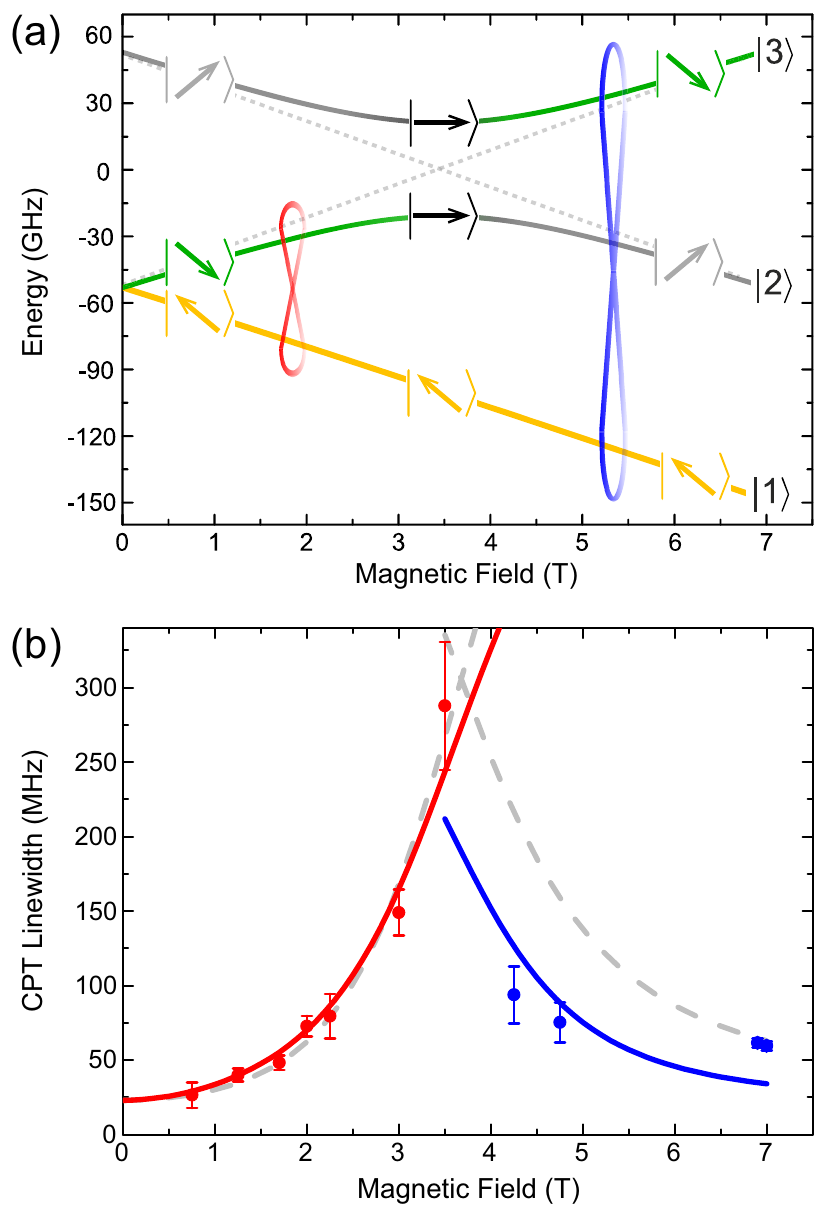}\\
\caption{(Color online) (a) Simulated ground states \cite{Supplement}, illustrating the spin state for magnetic field values below, above and at the avoided crossing (3.5\,T). 
(b) Full width at half maximum of the CPT dip as a function of the magnetic field, using a Lorentzian fit. Filled circles denote measured widths (for each transition, laser powers equal to four times the saturation power), with the error bars being the standard deviation of multiple measurements. The solid lines display the spin overlap (grey dashed lines) between states used for CPT, multiplied by a Boltzmann factor \cite{Supplement}. 
In panels (a) and (b), the color red (blue) indicates CPT realized between states $|1\rangle$ and $|2\rangle$ for $B < 3.5$\,T (between $|1\rangle$ and $|3\rangle$ for $B > 3.5$\,T).}
\end{figure}

In this work, we verified the presence of a spin in the ground state of the SiV$^{\text{-}}$ center in diamond, and probed its coherence using CPT. The ability to generate a coherent superposition of the SiV$^{\text{-}}$ spin state relying solely on optical fields establishes the route to full quantum control of the SiV$^{\text{-}}$ spin with picosecond operation timescale. This also allows for the implementation of all-optical dynamical decoupling schemes, enabling to further extend the coherence time of the spin state. The combination of ultrafast coherent control of individual spins and the high quality and reproducibility of the optical spectrum across multiple SiV$^{\text{-}}$ centers can serve to realize the basic components of a distributed quantum network.

\begin{acknowledgments}
We gratefully acknowledge financial support by the University of Cambridge, the European Research Council (FP7/2007-2013)/ERC Grant agreement no. 209636, FP7 Marie Curie Initial Training Network S$^3$NANO. This research has been partially funded by the European Community's Seventh Framework Programme (FP7/2007-2013) under Grant Agreement N$^\circ$611143 (\mbox{DIADEMS}). AIT and TG thank EPSRC Programme Grant EP/J007544/1 for support. Ion implantation was performed at and supported by RUBION, central unit of the Ruhr-Universit\"at Bochum. We thank D. Rogalla for the implantation, C. Pauly for SIL fabrication, the Sheffield group for their hospitality during the vector magnet measurements, A. Lenhard for the Monte-Carlo simulation of the intensity autocorrelation, as well as J. Barnes, J. Hansom, H. S. Knowles, J. M. Taylor, J. Eschner and G. Morigi for helpful discussions.
\end{acknowledgments}


\begin{thebibliography}{31}
\bibitem[{Sca()}]{ScalFact}
\bibinfo{note}{A scaling factor and an offset, common to both sides of the
  curve, are used for display along with the CPT width.}
\end{thebibliography}
\end{document}